\documentclass[12pt,preprint]{aastex}

\shorttitle{Radio Spectral Evolution of an Impulsive Flare}
\shortauthors{Bastian et al.}

\begin{document}

\title{Radio Spectral Evolution of an X-ray Poor Impulsive Solar Flare: \\
Implications for Plasma Heating and Electron Acceleration}
\author{T. S. Bastian}
\author{G. D. Fleishman\altaffilmark{1}}
\affil{National Radio Astronomy Observatory, Charlottesville, VA 22903}
\author{D. E. Gary}
\affil{New Jersey Institute of Technology, Newark, NJ}
\altaffiltext{1}{Ioffe Institute for Physics and Technology, St. Petersburg, Russia}

\begin{abstract}{\large }
We present radio and X-ray observations of an impulsive solar flare
that was moderately intense in microwaves, yet showed very meager
EUV and X-ray emission. The flare occurred on 2001 Oct 24 and was
well-observed at radio wavelengths by the Nobeyama Radioheliograph
(NoRH), the Nobeyama Radio Polarimeters (NoRP), and by the Owens
Valley Solar Array (OVSA). It was also observed in EUV and X-ray
wavelength bands by the TRACE, GOES, and {\sl Yohkoh} satellites. We
find that the impulsive onset of the radio emission is progressively
delayed with {\sl increasing} frequency relative to the onset of
hard X-ray emission. In contrast, the time of flux density maximum
is progressively delayed with {\sl decreasing} frequency. The decay
phase is independent of radio frequency. The simple source
morphology and the excellent spectral coverage at radio wavelengths
allowed us to employ a nonlinear $\chi^2$-minimization scheme to fit
the time series of radio spectra to a source model that accounts for
the observed radio emission in terms of gyrosynchrotron radiation
from MeV-energy electrons in a relatively dense thermal plasma. We
discuss plasma heating and electron acceleration in view of the
parametric trends implied by the model fitting. We suggest that
stochastic acceleration likely plays a role in accelerating the
radio-emitting electrons.
\end{abstract}


\section{Introduction}

Radio and X-ray radiation from solar flares have been used for many years to study energy release, plasma heating, and particle acceleration and transport. Joint studies of radio and X-ray radiation are particularly powerful because they probe populations of electrons at different energies, ranging from hot thermal electrons to nonthermal electrons of tens of keV emitting at soft X-ray (SXR) and hard X-ray (HXR) wavelengths, and nonthermal electrons with energies ranging from hundreds of keV to several MeV emitting $\gamma$-rays and centimeter- and
millimeter-wavelength radiation (e.g., Takakura et al. 1983; Klein et al. 1986; Bruggmann et al. 1994; Ramaty et al. 1994, Trottet et al. 1998, 2002).  Flares are conveniently divided into two broad classes, impulsive (or compact) and gradual (or extended) flares (Pallavicini et al. 1977) although more complex classification schemes have been proposed (e.g., Bai \& Sturrock 1989) in response to more detailed and comprehensive observations. Impulsive flares tend to occur in relatively high density, compact, coronal magnetic loops, and display a good correlation between the microwave and HXR emission (Kai et al. 1985, Kosugi et al. 1988). Gradual events tend to occur in lower density, larger scale coronal loops, emitting for significantly longer times. They tend to be ``microwave rich" (Bai \& Dennis 1985) by a factor of $\sim 5$ compared to the microwave-HXR correlation of impulsive events (Kosugi et al. 1988).  The spectral and temporal profiles of these two classes also follow a predictable evolution (e.g. Silva et al. 2000). Impulsive bursts show a soft-hard-soft spectral evolution in hard X-rays and a radio peak-frequency evolution that evolves from low to high, and often back to low (Melnikov et al. 2007). Gradual bursts, in contrast, show a soft-hard-harder evolution in X-rays and a radio spectral peak that is often constant or slowly increasing in frequency with time. Bursts that do not obey these general rules are often worthy of detailed study, especially when well observed, in order to test our understanding of the underlying physical mechanisms.

The impulsive flare of 2001 October 24 at 23:10 UT is an example
of such a burst, which was well observed by a broad suite of
instrumentation operating at radio and X-ray wavelengths and
therefore offers an opportunity to study a wide range of electron
energies. The event stands out paradoxically as an extremely X-ray-poor impulsive flare,
producing a moderately intense radio burst, yet very little EUV,
SXR, or HXR emission. It is similar in several respects to a radio
burst studied by White et al. (1992), an event for which available
radio data suggested that energy release occurred in a dense
plasma ($n_e\sim 5\times 10^{11}$ cm$^{-3}$). Veronig \& Brown
(2004) have identified a class of HXR sources that also occur in a
relatively dense plasma ($n_e\approx 1-2 \times 10^{11}$
cm$^{-3}$). As we show here, the Oct. 24 event is another example
of an event in which energy release occurs is a dense plasma.
However, unlike the events studied by Veronig \& Brown, the Oct.
24 event is highly impulsive; and unlike the event discussed by
White et al., the Oct. 24 event shows significant radio spectral
evolution. Although the physical conditions and the evolution of
the burst are unusual, we believe that the physical processes are
not. The excellent observational coverage at radio and X-ray
wavelengths provides us with the opportunity to address electron
acceleration in a dense plasma environment in some detail. In
particular, the excellent time-resolved radio spectroscopic data
available from OVSA and the NoRP has allowed us to fit the data to
a source model as a function of time and thereby deduce the time
variation of source parameters. These, in turn, have yielded
insights into the nature and consequences of energy release in a
dense plasma.

The paper is organized as follows:  In \S2 we present the radio,
EUV, and X-ray observations of the flare. In \S3 we argue that the
data may be understood in terms of fast electrons injected into a
relatively cool, dense magnetic loop. We explicitly fit a time
series of radio spectroscopic data to a simple model that embodies
these assumptions using a nonlinear $\chi^2$ minimization scheme. The parametric trends to emerge show that the
time series of spectra is well fit if the temperature of the
ambient plasma increases in time.  We discuss implications of the
model fits for plasma heating, electron acceleration, and electron
transport in \S4. We suggest that the lack of significant X-ray
emission can be understood if the magnetic mirror ratio is large,
thereby suppressing electron precipitation and reducing conduction
at the foot points. The relatively high plasma density in the
source renders it collisionally thick to electrons with energies
of some 10s of keV. We argue that collisional heating yields the
systematic plasma heating inferred from model fitting.
Furthermore, electrons are accelerated progressively to energies
exceeding MeV energies over over a time scale $<10$ s. We suggest
that stochastic processes may play a significant role in the
acceleration and transport of electrons. We conclude in \S5.

\section{Observations}

The flare on 2001 October 24 occurred in NOAA active region 9672 at
a heliocentric position of S18W13 from approximately 23:10-23:14 UT.
AR 9672 was in a $\beta\gamma\delta$ magnetic configuration and had
produced M and X class soft X-ray flares in previous days. While the
relatively strong radio event from AR9672 was noted by the NOAA
Space Environment Center, it was incorrectly associated with an
H$\alpha$ subflare that occurred nearly concurrently, from
23:07-23:12 UT, in AR 9678 (N06E33). No H$\alpha$ observations of AR
9672 are available at the precise time of the event. While no soft
X-ray emission was noted by NOAA/SEC from either the H$\alpha$
subflare in AR9678 or the radio burst in AR9672, weak
extreme-ultraviolet (EUV), soft X-ray (SXR), and hard X-ray (HXR)
emissions were in fact detected from the flare in AR9672 by the {\sl
Transition Region and Coronal Explorer} (TRACE; Handy et al. 1999),
the {\sl Yohkoh} Soft X-ray Telescope (SXT), and Hard X-ray
Telescope (HXT), respectively (Kosugi et al. 1992), as was a weak
SXR enhancement by the GOES 10 satellite. The radio burst was
well-observed in total intensity by the Owens Valley Solar Array
(OVSA; Gary \& Hurford 1994), the Nobeyama Radio Polarimeters
(NoRP), and by the Nobeyama Radioheliograph (NoRH; Nakajima et al.
1994).

Fig.~1a provides an overview of selected X-ray and radio
observations. The HXR and radio light curves show a simple, highly
impulsive rise and exponential decay over a duration of
approximately 4 min whereas the GOES SXR bands decay to half their
peak flux on a time scale $<10$ min. The horizontal line at the top
of Fig.~1a indicates the time range shown in Fig.~1b. Here, the HXR
light curve is shown with the NoRP frequencies. We now discuss the
EUV, X-ray, and radio observations in greater detail.

\subsection{General Comparison of NoRP and HXT L band Events}

To place the 2001 October 24 event in the  broader context of events
jointly observed in the radio and HXR bands,  Fig.~2 shows a scatter
plot of the 135 flares observed jointly by the NoRP at 17 GHz and
the HXT L band during the lifetime of the {\sl Yohkoh} mission.
Analogous figures have compared the 17 GHz peak flux with the HXR
counts from the hard X-ray spectrometer on board  {\sl Hinotori}
(Kai et al. 1985) and from the HXRBS experiment on board the {\sl
Solar Maximum Mission} (Kosugi et al. 1988). As in these previous
examples,  there is a clear correlation between the peak 17 GHz flux
and the peak HXT L band count rate. The Spearman (rank) correlation
coefficient is 0.57, and the two-sided significance of the
correlation -- i.e., the probability that the correlation
coefficient is larger than this value in the null hypothesis -- is
essentially zero, indicating the correlation is highly significant.
The flare of 2001 October 24 has a peak flux of 425 SFU\footnote{1
solar flux unit = $10^{-22}$ W m$^{-2}$ Hz$^{-1}$} at 17 GHz. The
regression curve implies that an HXT~L band count rate of more than
200 cts s$^{-1}$ sc$^{-1}$ is expected, yet a peak count rate of
only 7 cts s$^{-1}$ sc$^{-1}$ is measured (including a background of
$\approx 1$ ct s$^{-1}$ sc$^{-1}$). The flare is X-ray poor by a factor of $\sim 30$ relative to the regression curve.  We use the term ``X-ray poor" rather than ``microwave rich" because, as noted in the Introduction, the latter term is associated with gradual events showing excess microwave emission (Bai \& Dennis 1985, Cliver et al.
1986, Kosugi et al. 1988).

\subsection{EUV and X-ray Observations of the Oct 24 Event}

While the Oct 24 event is very X-ray poor it was nevertheless
detected by several instruments operating at EUV and X-ray
wavelengths. It was imaged by TRACE in the 171 \AA\ band with a
cadence of $\approx\!40$~s and an angular resolution of $1"$. An
image was obtained just prior to the impulsive rise at 23:10:46~UT
and another was obtained at 23:11:26 UT (Fig~1b). A difference image
formed from the latter image and one obtained at 23:10:05 UT reveals
discrete kernels of emission (Fig.~3a).

In soft X-rays, the GOES 10 background level was approximately C2.3
and the enhancement above background was approximately B2. The X-ray
emission was too weak to trigger flare mode observing by the SXT. As
a result, only one low-resolution (4.92") full-disk SXR image (AlMg
filter) is available during this brief impulsive flare, at 23:11:46
UT (Fig~1b, Fig.~3b).

The {\sl Yohkoh} HXT typically observed flares in four photon energy
bands --  L (13.9-22.7 keV), M1 (22.7-32.7 keV), M2 (32.7-52.7 keV),
and H (52.7-92.8 keV). In this case, however, the HXT flare mode was
not triggered and counts were therefore recorded in L band only. The
L band light curve (Fig.~1) shows a simple impulsive spike with a
maximum background-subtracted count rate of only 6 cts s$^{-1}$
sc$^{-1}$ and a FWHM duration of just 18 s. Integrating over the
full duration of the event yields sufficient counts to image the HXR
source with an angular resolution of $\approx\!5"$ (Fig. 3c).

The {\sl Yohkoh} Wide-band Spectrometer (WBS; Yoshimori et al. 1992)
may have made a marginal detection in energy channels up to 50 keV
(H. Hudson, private communication) but given uncertainties in the
calibration the implications for the HXR photon spectrum are highly
uncertain. We therefore do not consider WBS data here. 

Each of the maps shown in Fig.~3 is compared with a magnetogram. The
flare is associated with a magnetic bipole just south of the
dominant sunspot group in AR 9672.  The length of the source is
$\approx\!30"$ in SXR and its width is $\approx\!12"$. The SXR and
HXR sources are coincident (cf. Veronig \& Brown 2004) and are
consistent with a magnetic loop or unresolved loops, the latter
suggested by the discrete kernels seen in the TRACE 171 \AA \
emissions, presumed here to be footpoint emission.

\subsection{Radio Observations of the Oct 24 Event}

Radio imaging observations of the flare were obtained by the NoRH at
17 and 34 GHz. The NoRH observations were made early in the day,
local time, and the angular resolution was therefore reduced in one
dimension, yielding an elliptical beam of approximately $24"\times
10"$ at 17 GHz, and $13"\times 5"$ at 34 GHz, both with a position
angle $\phi\approx -40^\circ$. Figs.~4a and 4c show the 17 and 34
GHz maps in total intensity (Stokes I) overlaid on the magnetogram.
The source morphology evolves very little at radio wavelengths as a
function of time; we therefore show maps near the time of maximum
intensity. The peak brightness temperature of the 17 GHz source is
$4.6\times 10^7$ K whereas that of the 34 GHz source is $1.8\times
10^7$ K. Fig.~4b shows the 17 GHz map of the circularly polarized
flux (Stokes V parameter) superposed on the magnetogram,
demonstrating that it is polarized in the sense of the x-mode with
right-circularly polarized emission associated with positive
magnetic polarity. The degree of circular polarization (I/V) ranges
from 20\% left-circularly polarized (LCP; negative values of Stokes
V) to 10\% right-circularly polarized (RCP; positive values of
Stokes V). The observed change in the sense of circular polarization
along the length of the loop from RCP to LCP reflects the reversal
of magnetic polarity while the asymmetry in the degree of RCP/LCP
polarization indicates an asymmetry in the magnitude of the positive
and negative magnetic polarity.  The sources observed in the SXR,
HXR, and radio bands are coincident and quite similar in morphology.
All imaging data are consistent with the illumination of a simple
magnetic loop, or unresolved bundle of loops, straddling a magnetic
neutral line. We note that the loop is not visible in the NoRH 17
and 34 GHz maps as a discrete feature in the active region prior to
the flare. We estimate the 17 GHz brightness temperature prior to
the flare is $\sim\!10^5$ K.

The NoRP measures the total solar flux in Stokes I and V at
frequencies $\nu=1$, 2, 3.75, 9.4, 17, 35, and 80 GHz. Referring to
the NoRP observations summarized in Fig.~1b, we note several
features. First, the onset of the 35 GHz emission is delayed by
$\approx\!3$ s relative to that at 17 GHz which is, in turn, delayed
by $\approx\!3$ s relative to the onset of the HXT L band emission.
Second, the time of the 17 GHz flux maximum is delayed by
$\approx\!3$ s relative to that at 35 GHz and the maximum of the 9.4
GHz light curve is delayed by 12 s relative to that at 17 GHz. Note,
however, that a ``shoulder" in the 9.4 GHz emission roughly coincides
with the 17 and 35 GHz maxima, a point discussed further in \S4. Third, the decay profiles of the 9.4,
17, and 35 GHz light curves are nearly identical and are exponential
after approximately 23:11:20 UT with a decay time of $\approx\!48$
s. Fourth, there is no detectable radio emission at 3.75 GHz.
Finally, while the flare was detected at 80 GHz the signal is very
noisy (4-5$\sigma$). We note that both the sensitivity and the
calibration of the 80 GHz flux on this date are rather poor, the
absolute flux calibration being no better than $\approx 40\%$ (H.
Nakajima, private communication).

These features are confirmed and extended by the OVSA observations.
On 2001 October 24, OVSA employed a high cadence observing mode to
produce 24 point spectra between 1.2-14.8 GHz every 2 s in total
intensity. A detail of the impulsive onset of the flare is shown in
Fig.~5a which plots the HXT L band data, three OVSA frequencies
(11.2, 12.4, and 14.8 GHz, labeled with a prefix ``O"), and  three
NoRP frequencies (9.4, 17, and 35 GHz, labeled with a prefix ``N")
for a period of 20~s. The frequencies 12.4, 14.8, 17, and 35 GHz
show a progressive delay in their flux onset relative to the HXT L
band emission. The onset of emission at  9.4 and 11.2 GHz show the
opposite pattern, however.  These trends are also clearly seen in
the dynamic spectrum of the OVSA data (Fig.~6). For frequencies
$\gtrsim\!12$ GHz the flux onset is increasingly delayed with
frequency whereas the opposite trend is seen for $\nu\lesssim\!12$
GHz.

Fig.~5b shows a detail of the decay phase for the same selection of frequencies shown in Fig.~5a. Each frequency has been scaled so that for times later than 23:11:20~UT the light curves are overlaid, showing that the exponential decay from 9.4 to 35 GHz is essentially identical after 23:11:20~UT. Fig.~5b also illustrates the progressive delay of the flux maximum with decreasing frequency. This feature is more clearly illustrated in Fig.~6, in which it is readily seen (dashed line) that the time of the flux maximum increases with decreasing frequency. There is no detectable emission below $\sim\!5-6$ GHz.

The radio spectral evolution of the flare is summarized in Fig.~7,
which shows composite spectra using OVSA data (plus symbols) and NoRP data (diamonds) during
the impulsive rise and exponential decay. The spectral maximum is
near 17 GHz during the rise phase but declines to $\sim\!10$ GHz
during the decay phase.

\subsection{Observational Summary}

Given the rather extensive data set it is useful to consolidate and
summarize our findings regarding the flare on 2001 October 24:

\begin{itemize}
\item{The flare is one of the most X-ray poor events of those jointly observed by the NoRP and {\sl Yohkoh} HXT. }
\item{The NoRH 17 and 34 GHz images, and the {\sl Yohkoh} SXT and HXT L band images are all coincident and similar in morphology. The source is consistent with a simple magnetic loop or unresolved bundle of loops. The TRACE 171\AA\   difference image reveals small kernels consistent with loop foot points.}
\item{Due to the paucity of X-rays, the photon spectrum was not measured and the electron spectrum is therefore unconstrained by HXR emission. }
\item {The observed radio emission is relatively strong and the spectral coverage is excellent, allowing the following observations to be made: }
\begin{itemize}
\item{Radio emission is detected up to 80 GHz. The spectral maximum is $\sim\!17$  GHz during the impulsive rise, declining to $\sim\!10$ GHz during its decay. Radio emission is undetectable below $\sim\!5-6$ GHz.}
\item{The impulsive onset of emission shows a progressive delay with {\sl increasing} frequency for $\nu\gtrsim 12$ GHz; the trend reverses below $\approx\!12$ GHz.}
\item{In contrast, the time of absolute flux maximum shows a progressive delay with {\sl decreasing} frequency for all frequencies measured.}
\item{For times later than approximately 23:11:20~UT the decay profile of the radio emission is independent of frequency over two octaves and is characterized by a decay time of $\approx 48$ s.}
\end{itemize}
\end{itemize}

\noindent Despite the relatively simple source morphology in the
X-ray and radio bands, and the single impulsive spike shown in the
HXR and radio light curves, this event displays considerable
spectral richness at radio wavelengths. In the next section we
develop a source model and perform parametric fits to the time
series of radio spectra. The parametric trends to emerge from these
fits then guide our physical interpretation of the event in \S4.

\section{Model of the Radio Source}

We now direct our attention to the well-observed radio source and
fit a simple model to the radio data as a function of time. A fully
self-consistent, three dimensional, time-dependent, inhomogeneous
model of the source is beyond the scope of this work. We instead
account for the essential features of the source in terms of a
magnetized volume containing thermal plasma and energetic electrons.
Similar models have been developed to account for the spectral
(e.g., Klein 1987; Benka \& Holman 1992; Bruggmann et al. 1994),
temporal (e.g., Bruggmann et al. 1994), or spatial (Bastian et al.
1998; Nindos et al. 2000; Kundu et al. 2004) properties of radio
observations. Typically, these efforts have been strongly
under-constrained in the temporal, spatial, and/or spectral domains.
Moreover, model fitting has typically been limited to adjusting
model parameters until a reasonable fit by eye is obtained.

In the present case, we have comparatively good spectral and
temporal coverage of the event at radio wavelengths. While imaging
data are limited, the source is apparently rather simple.  It is
therefore possible to make simplifying assumptions regarding the
source volume and to exploit nonlinear optimization techniques to
fit the observed radio spectrum. Moreover, such fits can be
performed on the time series of spectra, from which the time
variation of fitted parameters can be deduced. As such, the fitting represents a means of time-resolved radio-spectral inversion. While the model is
admittedly crude, it nevertheless allows us to identify trends in
the time variation of critical parameters in an unbiased fashion. In
other words, rather than postulating a time dependence for physical
parameters of interest, the time dependence emerges from the model
fits.  In this section we describe the model and the underlying
assumptions, explore the parametric dependencies of the model, and
then present the fitting scheme and the results derived from the
fits in detail.

\subsection{Model and Assumptions}

We assume the observed radio emission is the result of
gyrosynchrotron radiation from energetic electrons in a magnetic
loop. A striking feature of the radio spectra is the relatively
steep roll-off in the emission below $10-17$ GHz, with no emission
detectable at all below $\approx 6$ GHz, a feature that persists for
the duration of the event. The spectrum of the radio emission is
clearly nonthermal, yet the brightness temperature of the 17 and 34
GHz sources is unexceptional, characteristic of optically thin
emission. We therefore suggest that the Razin effect (Razin 1960,
Ginzburg \& Syrovatskii 1969) plays a role in this flare. The Razin
effect  suppresses gyrosynchrotron emission in the presence of an
ambient plasma below a cutoff frequency
$\nu_R\approx\nu_{pe}^2/\nu_{Be}\approx 20n_{th}/B$, where $n_{th}$
is the thermal electron density and $B$ is the source magnetic
field. For $\nu_R\sim 10-17$ GHz and and an ambient density 
$n_{th}\sim 10^{11}$ cm$^{-3}$, $B=120-200$ G. For a given cutoff frequency
$\nu_R$, higher magnetic field strengths require higher thermal
electron densities.

The thermal free-free absorption coefficient at radio wavelengths is
$\kappa_{ff}\approx 0.2n_e^2 T^{-3/2}\nu^{-2}$ for coronal
conditions (Dulk 1985). The optical depth $\tau_{ff}$ through a
plasma layer $L$ is $\tau_{ff}=\kappa_{ff}L$. Substituting plausible
values for the relevant parameters -- $T=2\times 10^6$ K, $L=5\times
10^8$ cm, and $n_{th}=10^{11}$ cm$^{-3}$ -- we find that
$\tau_{ff}\approx 350\nu_9^{-2}$, where $\nu_9$ is the frequency in
GHz. The optical depth is even larger for cooler plasma. It is clear
that if Razin suppression is important, free-free absorption is,
too. Consequently, we assume that Razin suppression
and free-free absorption each play a role in determining the shape of
the radio spectrum and include both in our model calculations (cf.
Ramaty \& Petrosian 1972, Klein 1987).

As previously noted, the source morphology changes little during the
event. We therefore consider a homogeneous source with an area
$A=2\times 10^{18}$ cm$^2$ ($12"\times 30"$) and a depth $L=9\times
10^8$ cm (12"), consistent with the X-ray and radio imaging (see
\S2.4). The source volume is assumed to contain thermal plasma with
a density $n_{th}$ and a temperature $T$, permeated by a coronal
magnetic field $B$ with an angle $\theta$ relative to the line of
sight. Finally, it is assumed that an isotropic, power-law
distribution of energetic electrons $N(E)dE=KE^{-\delta}dE$ is
uniformly mixed with the thermal plasma, with a normalization energy
$E_o$ and a high-energy cutoff energy $E_c$. The total number
density of nonthermal electrons between $E_o$ and $E_c$ is $n_{rl}$.
No electron transport effects are explicitly included in the model calculation.

The computation of radio emission from even as simple a source model
as this depends on many parameters. To make further progress,
additional assumptions are necessary. First, we assume the magnetic
field remains constant for the duration of the event. However, as a
means of estimating the dependence of the fitted parameters on the
magnetic field, different values of the magnetic field were assumed
for different fitting runs. Second we assume the number density of
the background thermal plasma also remains constant. In fact, as
discussed further in \S3.3, preliminary fits allowed $n_{th}$ to
vary but the fitted values showed little variation with time. Hence,
it was typically held fixed. An additional justification for this
second assumption is that, given the weak EUV and HXR emission,
there was little deposition of energy near the loop footpoints and
hence, little evaporation of plasma into the coronal loop. The
incremental increase in density due to evaporation is therefore
presumed to be small. We do allow the temperature of the thermal
background plasma to vary with time, however. Finally, we note that
the radio emission is insensitive to $E_o$ and therefore set it
equal to 100 keV.

To summarize, we model the radio source as a simple homogenous
volume with an area $A$ and a depth $L$. The volume is filled with a
thermal plasma with a number density $n_{th}$ and is permeated by a
magnetic field $B$ oriented at an angle $\theta$ to the line of
sight; $n_{th}$, $B$, and $\theta$ are assumed to be constant during
the flare but the plasma temperature is allowed to vary. A power-law
distribution of $n_{rl}$ energetic electrons per cubic centimeter
between energies $E_\circ$ and $E_c$ is uniformly mixed with the
background plasma. The electron pitch angle distribution is assumed
to be isotropic. Typically two or three out of four parameters --
$n_{rl}$, $T$, $E_c$, and $\delta$ -- were allowed to vary in
fitting the data to the model.

\subsection{Model Dependencies}

It is useful to make a brief aside at this point to outline the
computation of the radio emission spectrum and to elucidate the
dependencies of the computed spectrum on various parameters of
interest. For the homogeneous source described above, the source
flux at a given frequency is computed as (Ramaty \& Petrosian 1972;
Benka \& Holman 1992)

\begin{equation}
S(\nu, B, \theta, T, n_{th}, E_o, E_c, n_{rl}, \delta)= 2 k_B
{\nu^2\over c^2} {1\over{\mu^2(\nu,n_{th},B,\theta)}}{j_T \over
\kappa_T}{A\over D^2}[1-\exp{(\kappa_TL)}]
\end{equation}

\noindent where $\mu(\nu,n_{th},B,\theta)$ is the index of
refraction, $j_T = j_{ff}(\nu, T, n_{th})+j_{gs}(\nu,B,
\theta,n_{th},n_{rl},E_o, E_c,\delta)$ is the total emissivity due
to the free-free and gyrosynchrotron mechanisms, $\kappa_T =
\kappa_{ff}(\nu,T,n_{th})+\kappa_{gs}(\nu,B,\theta,n_{rl},E_o,E_c,\delta)$
is the total absorption coefficient, $A/D^2$ is the solid angle
subtended by the source, and $D$ is the distance from the observer
to the source (1 AU).

The thermal free-free emission and absorption coefficients used for
the computation of the source function are readily calculated (e.g.,
Lang 1980); those for gyrosynchrotron emission and absorption are
more problematic. The expressions for the emissivity and absorption
coefficient given by Ramaty (1969), or variants thereof (e.g., Benka
\& Holman 1992), are cumbersome and their calculation is
computationally intensive. For the iterative $\chi^2$ minimization
scheme we employ, it is far more efficient to use the approximate
expressions derived by Klein (1987) which include the effects of
Razin suppression and are quite accurate for radio frequencies
greater than the first few harmonics of the electron gyrofrequency
($\nu\gg \nu_B=eB/2\pi m_ec$) and for values of $\theta$ that are
not too small ($>30^\circ$).

Fig.~8 shows model spectra calculated for various sets of
parameters. All four panels show a reference spectrum (solid line)
due to gyrosynchrotron emission from electrons {\sl in vacuo}
($n_{th}=0$). The parameters used for the reference spectrum are:
$B=150$ G, $\theta=60^\circ$, $n_{rl}=5\times 10^6$ cm$^{-3}$,
$E_\circ=100$ keV, $E_c=2.5$ MeV, $\delta=3$, $A=2\times 10^{18}$
cm$^{2}$, and $L=9\times 10^8$ cm.  In Fig.~8a the dashed lines
indicate spectra where the source parameters are identical to those
used to compute the reference spectrum except that an ambient
thermal plasma is present in the source. The ambient plasma has a
temperature of $2\times 10^6$ K and its density varies from
$n_{th}=2\times 10^{10}$ cm$^{-3}$ to $5\times 10^{11}$ cm$^{-3}$.
The combined action of free-free absorption and Razin suppression
strongly reduces the radio emission below 10-20 GHz for values of
$n_{th}>5\times 10^{10}$ cm$^{-3}$. Moreover, as the density
increases, the high-frequency emission is enhanced. This enhancement
is entirely due to the thermal free-free contribution.

In Fig.~8b, the temperature is again $T=2\times 10^6$ K and the
ambient density is held fixed at $n_{th}=10^{11}$ cm$^{-3}$. Here,
the magnetic field strength is allowed to vary from $B=150$ to 350
G. All other source parameters are identical to those used to
compute the reference spectrum. Note that, in contrast to
gyrosynchrotron emission in which Razin suppression plays no role,
the spectral maximum is insensitive to $B$ although the peak flux
increases with $B$.

In Fig.~8c, the density is again held fixed at $n_{th}=10^{11}$
cm$^{-3}$, $B=150$ G, but the plasma temperature now varies from
$T=10^6$ to $2\times 10^7$ K. All other parameters are again the
same as those used to compute the reference spectrum. For cool
plasma, the combination of Razin suppression and thermal absorption
strongly reduces the microwave emission. As the temperature of the
ambient plasma increases, all other parameters held fixed, the
microwave emission increases at lower frequencies because the
free-free absorption drops as the temperature increases, but
decreases at the highest frequencies owing to the corresponding
decrease of the optically thin free-free opacity, and the spectral
maximum moves toward lower frequencies. At high enough temperatures
the plasma becomes optically thin to free-free absorption ($\sim
10^7$ K in this example) and the spectrum becomes insensitive to
further increases in temperature except the extreme low-frequency
end of the of the spectrum where the emission is due to optically
thick thermal radiation of the plasma and is proportional to its
temperature.

Finally, in Fig.~8d, the thermal plasma density is held fixed
($n_{th}=10^{11}$ cm$^{-3}$), $T=2\times 10^6$ K, $B=150$ G, and the
number of energetic electrons $n_{rl}$ between $E_o$ and $E_c$,
varies from $10^6$ to $2\times 10^7$ cm$^{-3}$. All other parameters
are the same as those used to compute the reference spectrum. The
microwave flux varies with $n_{rl}$ but the spectral maximum is
insensitive to variations in $n_{rl}$. This point has also been made
by Belkora et al. (1997), Fleishman \& Melnikov (2003), and Melnikov
et al. (2007). The optically thin slope steepens with increasing
$n_{rl}$ as the influence of the dense thermal plasma becomes
relatively less important. At low values of $n_{rl}$ the
contribution of free-free emission is significant and acts to
flatten the spectrum.

While the source model is relatively simple, it yields a wide
variety of spectral shapes which, in turn, yield insight to the
source parameters. We now turn to fitting this simple source model
to the observed data.

\subsection{Model Fitting and Results}

The excellent radio spectroscopic and temporal coverage of this
event allow us to fit the observed radio  spectra as a function of
time. Lee \& Gary (2000) used a similar approach to fit an electron
transport model to OVSA data although a different "goodness of fit"
parameter was optimized. Here, we have developed a nonlinear code
that adjusts model parameters to minimize the $\chi^2$ statistic
using the downhill simplex method (Press et al. 1986). The $\chi^2$
statistic is  computed as

\begin{equation}
\chi^2 = {1\over {N-n}}\sum_{i=1}^N {{(S_i^{obs}-S_i^{mod})^2}\over{\sigma_i^2}}
\end{equation}

\noindent where $S_i^{obs}$ is the flux density at frequency $i$
observed by either OVSA or the NoRP, $S_i^{mod}$ is the
corresponding model value, $\sigma_i^2$ is the estimated variance of
the observed value, $N$ is the number of data points, and $n$ is the
number of free parameters. The signal variance $\sigma_i^2$ at each
frequency is empirically known prior to the flare. It is the result
of receiver noise, the sky (35 and 80 GHz), and the Sun itself.
Radio emission from the flare makes an additional contribution that
can be estimated from the known intensity of the signals and the
properties of the instrument.

We have fit the model to composite radio spectra observed by OVSA
(5-14.8 GHz) and the NoRP (17, 35, and 80 GHz). Data points below 5
GHz are ignored because they contain no signal. Inspection of the
spectra indicates a probable systematic calibration error between
the two instruments with a magnitude of $\approx\!10-20\%$ (Fig.~8).
The time resolution of the NoRP data points were averaged to 2 s to
match the time resolution of the OVSA data. A total of 25 spectra
were fit to data sampled between 23:10:54 and 23:11:44 UT, as
indicated in Fig.~1b. The spectra at earlier and later times had
insufficient signal-to-noise ratios at too many points to yield
reliable fits.

While the spectral coverage of this event at radio frequencies is
excellent, it is nevertheless insufficient to allow unique sets of
parameters to be established from fits to the data. Excluding
frequencies below 5 GHz, we fit to only sixteen data points for each
time average. Indeed, as explained in \S3.1, many of the source
parameters are held constant and only two or three out of four
parameters were varied. Clearly, the models resulting from this
procedure, while providing good fits to the data, are non-unique. We
set those parameters that remain constant for all fits to values we
judge to be reasonable, but are fully aware that other, similar,
values may in fact yield fits that are consistent with the data. We
therefore consider the fits and the trends that emerge from the time
series of fitted parameters to be qualitative guides in our physical
interpretation of the flare in \S4.

Fig.~9 shows the results of two- and three-parameter fits to the 25
spectra observed from 23:10:54 to 23:11:44 UT. In the two-parameter
fit (solid line), only the number of fast electrons, $n_{rl}$, and
the plasma temperature, $T$, were allowed to vary and the spectral
index and cutoff energy were held fixed, with $\delta=3.5$  and
$E_c=2.5$ MeV, respectively, as were all other relevant parameters:
$B=165$ G, $\theta=60^\circ$, $E_\circ=100$ keV, $A=2\times 10^{18}$
cm$^{2}$, and $L=9\times 10^8$ cm.   The value for the spectral
index is entirely consistent with previous findings (e.g., Ramaty et
al. 1994, Trottet et al. 1998, 2000; L\"uthi et al. 2004). Three
parameter fits are also shown, the third parameter being either
$\delta$ (dashed line) or $E_c$ (dotted line). As discussed in
\S3.1, little improvement was seen in the (three-parameter) fits
when the thermal plasma density, $n_{th}$ was free to vary. The
fitted value of $n_{th}$ remained nearly constant ($n_{th}\approx
10^{11}$ cm$^{-3}$) in all cases and was therefore held fixed at
this value for the fits shown in Fig.~9. When $\delta$ was allowed
to vary it, too, remained relatively constant or in some cases
showed a tendency to smaller values; i.e., spectral hardening. When
$E_c$ was allowed to vary, best fit values during the rise phase of
the flare yielded $E_c\approx 2$ MeV whereas for times later than
approximately 23:11:12 UT, larger values of $E_c$ resulted ($E_c\ge
10$) MeV. We note that the fits become insensitive to the precise
value of $E_c$ when it exceeds $\sim\!10$ MeV because the high
frequency radio data most sensitive to $E_c$ are sparse and, in the
case of 80 GHz, poorly calibrated.

Fig.~10 shows a comparison between the fitted values for $n_{rl}$
and $T$ resulting from the two-parameter fits and the radio and HXR
emission as a function of time. We note that the three-parameter
fits yield similar results for the time variation of $n_{rl}$ and
$T$. The shaded area in each panel indicates the range of fitted
values that result from varying the magnetic field strength. The
time variation of $n_{rl}$ and $T$ are shown for three fixed values
of the magnetic field (150, 165, and 180 G).  Two points can be made
about the apparent trends: First, the total number of energetic
electrons tracks the variation of the high frequency (35~GHz) radio emission
for the interval that was fit quite well, as expected for optically thin emission.
Second, the temperature of the ambient
plasma increases with time. With $B=165$ G, $T$ increases from $\sim
2-3\times 10^6$ K to $\sim 8-10 \times 10^6$ K during the time range
considered. Both $T$ and $n_{rl}$ depend on $B$: the fitted value of
$n_{rl}$ decreases and that of $T$ increases with increasing $B$.
Referring to Fig.~8c, we note that the fits are relatively insensitive to plasma
temperature once $T$ approaches $10^7$ K (Fig.~10b) because the
optical depth to free-free absorption becomes negligible. We find no evidence that the observed spectral evolution requires multiple discrete injections of electrons characterized by different energy distribution functions, as suggested by, e.g., Trottet et al. (2002), for certain events. 

We checked the fitted temperature and the emission measure implied
by the assumed density (i.e., $EM\sim n_{th}^2AL$) against the GOES
SXR observations for consistency (White et al. 2005). A peak temperature near $10^7$ K is
consistent with the GOES 0.5-4 and 1-8 \AA\ data but the density
inferred from the GOES data is perhaps a factor of two smaller than
that assumed here. Given that the GOES event is very small and the
uncertainties in the SXR background are relatively large, and given
that the parametric fits depend on a number of assumptions, we
conclude that the consistency between the fits and the SXR data is
rather good.

\section{Discussion}

The ``standard model" of flares involves energy deposition in the
chromosphere by a flux of nonthermal electrons precipitating from
one or more magnetic loops, resulting in heating and ``evaporation"
of chromospheric plasma into the corona where it emits SXR
radiation. We suggest that a difference between this flare and the
``standard model" is that significant chromospheric evaporation did
not occur. As noted in the previous section, the spectral fits do
not require an increase in the thermal plasma density during the
course of the flare. Furthermore, there is a paucity of EUV and SXR
emission from hot thermal plasma, suggesting little increase in
emission measure of coronal plasma and hence, little increase
in the thermal plasma density.

Why is there so little chromospheric evaporation? One reason may be
that the flaring loop has a large (albeit asymmetric) magnetic
mirror ratio, defined as $m=B_{FP}/B_{LT}=\sin^{-2}\alpha_{lc}$, the
ratio of the magnetic field strength near the magnetic footpoint to
that at the loop top where the magnetic field is weakest, and
$\alpha_{lc}$ is the loss-cone angle. A large value of $m$ implies a
small value for $\alpha_{lc}$, little precipitation of energetic
electrons from the loop and hence, little EUV and HXR footpoint
emission. For example, a mirror ratio $m=5$ implies only
$\sim\!10\%$ of an isotropic particle distribution would precipitate
from the magnetic loop if magnetic trapping and weak scattering were
the only relevant transport processes (e.g., Melrose \& Brown 1976).
A second factor is that the coronal loop, with $n_{th}=10^{11}$
cm$^{-3}$, is collisionally thick to electrons with energies
$\lesssim 30$ keV (Veronig \& Brown 2004). Such electrons would
deposit the bulk of their energy in the ambient plasma rather than
at the loop footpoints. One might suppose that if the precipitation
of energetic electrons into the footpoints is limited, thermal
conduction might nevertheless drive evaporation. However, if the
mirror ratio is large the footpoint area is small and evaporation
due to conduction can also be strongly reduced. These considerations
are consistent with the observed morphology of the HXT L band
source shown in Fig.~3c.  The idea of a large mirror ratio
stands in contradiction to our simple model, for which a
magnetically homogeneous source was assumed. This may not be a
concern, however, if the magnetic loop is strongly convergent near
the chromosphere (Dowdy et al. 1985).

The time interval during which plasma heating is observed via
spectral fitting is from 23:10:50 to 23:11:20 UT, so that $\Delta
t\approx\!30$ s. During this time, the data suggest the temperature
of the ambient thermal plasma in the flaring source increased from
2-3 MK to 8-10 MK. We note that the source was presumably already
heated from $\sim 0.1$ MK to 2-3 MK prior to the time range for
which fits were computed. Whether $T$ continued to increase
significantly beyond 23:11:20 UT is not established because the
model fitting becomes insensitive to temperature when $T>10^7$ K
although the GOES SXR emission suggest that it did not. Since both
the radiative and conductive loss times are long compared to the
duration of the flare, the net change of thermal energy density
during the observed temperature increase is $\Delta E_{th}\approx
3n_{th}k_B\Delta T\approx 300$ ergs cm$^{-3}$, yielding a mean
energy deposition rate of $\dot E=\Delta E_{th}/\Delta t\approx 10$
ergs s$^{-1}$ cm$^{-3}$. The magnetic energy density
$u_B=B^2/8\pi\approx 1000$ erg cm$^{-3}$. Hence, the energy
deposition rate required to heat the plasma amounts to $\approx\!1\%$ of 
$u_B$ each second and a total energy deposition of $\sim\!5\times
10^{29}$ ergs.

What heats the plasma? One possibility, to which we have already
alluded, is energy loss via Coulomb collisions. By virtue of its
high density the coronal loop is collisionally thick to electrons
with energies $\lesssim\!30$ keV. Unfortunately, no HXR observations
of photon energies greater than the HXT L band are available because
the flare did not trigger the {\sl Yohkoh} flare mode. The spectral
index and low-energy cutoff of the distribution of HXR-emitting
electrons are unknown and hence, the number of electrons with
energies of 10s of keV, cannot be inferred from the HXR
observations. It may even be possible that thermal bremsstrahlung contributes to the HXT L band counts. However, if we suppose the mean energy of the
HXR-emitting electrons is $\approx 30$ keV, then $\approx 2\times
10^8$ such electrons are needed each second to account for the
plasma heating via collisions.

We now consider the radio-emitting electrons. The peak number
density of energetic electrons with energies $>100$ keV was found to
be $n_{rl}\approx 10^7$ cm$^{-3}$. Note that while $n_{rl}$ is
normalized to $E_\circ=100$ keV, for $B=165$ G the bulk of the
emission at radio frequencies between $\approx\!10-40$ GHz is in
fact contributed by electrons with energies of $\approx\!0.7-3$ MeV,
a result that has been previously noted (Ramaty et al. 1994, Trottet
et al. 1998, 2000; L\"uthi et al 2004). The energy contained in
these electrons is only $\sim\!0.1$ ergs cm$^{-3}$, an insignificant
fraction of that required to heat the ambient plasma. We conclude
that the radio-emitting electrons themselves play no role in heating
the plasma; but what if the power-law distribution of
radio-emitting electrons extends to lower energies? The collisional
energy deposition rate is

\begin{equation}
\label{Edot}
 \dot E\approx \int_{E_l}^{E_C} n(E)E/t_E\ dE,
\end{equation}

\noindent with

\begin{equation}
\label{t_E} t_E=170 (\gamma-1) n_{10}^{-1} \beta\approx 20 E_{100}^{3/2} n_{10}^{-1}~s,
\end{equation}

\noindent where the approximate expression for the lifetime $t_E$
applies to mildly relativistic electrons (Melrose \& Brown 1976).
Here, $\gamma=1/\sqrt{1-\beta^2}$ is the Lorentz factor,
$\beta=v/c$, $E_{100}$ is the electron energy in units of 100 keV,
and $n_{10}$ is the thermal plasma density in units of $10^{10}$
cm$^{-3}$. Taking $\delta=3.5$ (\S3) we find $E_1\approx 40$ keV
yields an energy deposition rate of 10 ergs cm$^{-3}$ s$^{-1}$, in
agreement with the mean energy deposition rate, and the total number
of electrons $n>40$ keV is found to be $\approx\!10^8$ cm$^{-3}$
s$^{-1}$. The inferred electron energy and number density are similar
to those needed to account for plasma heating by Coulomb
collisions. We conclude that the population of electrons responsible
for the HXT L band emission and plasma heating can be regarded as
part of the same power-law distribution of electrons responsible for
the radio emission.

While this picture is appealing, it is incomplete. Thus far we have
based our discussion on the results of the two-parameter ($n_{rl},\
T$) fits to the time series of radio spectroscopic data for which
the electron energy distribution is  assumed to be a power-law
described by fixed values of $E_\circ$, $E_c$, and $\delta$. While
there are hints in the three-parameter fits of possible evolution of
$\delta$ and/or $E_c$, such variations did not yield significantly
better fits during the time interval considered. In \S2.3, however,
we considered the impulsive onset of the event and found that onset
time progressively increased with frequency for $\nu \gtrsim 12$ GHz
(Fig.~5a).  The mean energy of the emitting electrons
$\bar{E}\propto (\nu/\nu_B)^{0.5+0.085\delta}$ (Dulk 1985); hence,
higher radio frequencies are emitted by higher energy electrons for
a fixed magnetic field. The progressive delay in the onset of
impulsive emission for $\nu\gtrsim 12$ GHz suggests that the more
energetic the electrons, the later they appeared in the source. The
onset times of emission for $\nu\lesssim 12$ GHz may at first seem
inconsistent with the trend seen at higher frequencies. However,
they can be understood in terms of the overall pattern of strong
free-free absorption diminishing with time as the plasma temperature
increases. Consider 9.4 GHz: at the time of flux onset the 9.4 GHz
flux is strongly reduced by free-free absorption and Razin
suppression. As the plasma heats the free-free opacity declines and
the gyrosynchrotron emission increases. The triangle at 23:11:02 UT
in Fig.~1b, and the dotted line in Fig.~7, indicates the time of the
35 GHz flux maximum. In Fig.~1b it clearly coincides with the
"shoulder" seen in the 9.4 GHz light curve, a feature that roughly
coincides with the maximum of $n_{rl}$. It is not until the plasma
heats still further that the free-free opacity becomes negligible at
9.4 GHz and the flux maximum is achieved, well after the maximum of
$n_{rl}$.

Therefore, despite the complicating factor of free-free absorption,
we conclude that the timing delays seen in the onset of  the radio
emission relative to the HXT L band, and from lower to higher
frequencies, are consistent with progressive acceleration of
electrons with time during the impulsive rise. The net time elapsed
from the acceleration of electrons with energies of 10s of keV to
those with energies of 100s of keV is $\sim 2-3$ s, and from 100s of
keV to MeV energies  is $\sim4-6$ s (Fig.~5a).  After the initial
energization to MeV energies, there does not appear to be any
further significant evolution of the form electron energy spectrum,
as determined by spectral fitting. We return to this point below.

Finally, we consider the observed frequency independence of the
decay of the radio emission of this event, a feature shared with the
flare described by White et al. (1992). This feature is clearly at
odds with flare transport models in which the weak diffusion regime
and magnetic trapping play a significant role (e.g., Bruggmann et
al. 1994; Bastian et al. 1998; Lee et al. 2000; Kundu et al. 2001). If weak diffusion and 
trapping were relevant, higher
energy electrons would be lost from the distribution more slowly than
lower energy electrons because the collision frequency $\nu_c=1/t_E$ and the energy loss rate
decrease with energy according to equation (\ref{Edot}).
Consequently, higher frequency radio
emission would decay more slowly than lower frequency emission (see Kundu
et al. 2001 for a vivid example). Yet, the radio decay profiles for
the Oct 24 event are essentially identical (Fig.~5b); they are exponential with a decay time
of 48~s for all frequencies with detectable emission for times
later than approximately 23:11:20 UT. We conclude that while Coulomb
collisions play a dominant role in the energy loss of HXR-emitting
electrons (10s of keV), they do not appear to play a role in the
transport of the MeV radio-emitting electrons.  Instead,
wave-particle interactions mediate electron transport.

To summarize our findings, electron acceleration occurs during the
impulsive rise phase of the flare, progressing from energies of 10s
of keV to more than an MeV in $\lesssim 10$ s. There is no
compelling evidence for further significant evolution of the
electron spectrum during the  remainder of the flare based on model
fitting. The flare plasma is collisionally thick to electrons with
energies of some 10s of keV; these electrons heat the plasma. The
increase in plasma temperature leads to a frequency-dependent
decrease in free-free opacity, thereby explaining the increasing
delay of the radio flux maximum with decreasing frequency.  The
decay phase of the flare at radio frequencies is incompatible with
Coulomb collisions; we conclude that wave-particle interactions
mediate electron transport.

Given that wave-particle interactions dominate electron transport at MeV energies, we briefly
consider the role of stochastic processes for electron acceleration. Possibilities for the relevant wave modes include
whistler waves (Hamilton \& Petrosian 1992, Pryadko \& Petrosian
1997, Petrosian \& Liu 2004), fast-mode MHD waves (Miller et al.
1996), or others (see Toptygin 1985, Miller et al. 1997, for
reviews). To be more concrete, consider a turbulent spectrum of
whistler waves with a total energy density $u_W$. Hamilton \&
Petrosian (1992) have considered stochastic heating and acceleration
by whistler waves, including the effects of Coulomb collisions.
Assuming the spectral energy density of whistler wave turbulence is
isotropic and can be described by a power law dependence $\propto
k^{-q}$ for wave numbers $k>k_\circ$, the average pitch angle and
momentum diffusion coefficients, $\langle D_{\alpha\alpha}\rangle$
and $\langle D_{pp}\rangle$, are well known (e.g., Melrose 1980,
Eqns. 13 in Hamilton \& Petrosian 1992). The electron mean free path
is given by Hamilton \& Petrosian as $\lambda={{\beta c}/{\langle
D_{\alpha\alpha}\rangle}}\propto (\beta\gamma)^{2-q}$. For a source
of size $L$, taken here to be the observed loop length of $\approx
2\times 10^9$ cm, the lifetime in the source $t_s$ is given by
$t_s=L^2/\lambda\beta c\propto (\beta\gamma)^{q-2}/\beta$.
Expressing the energy density in the turbulent spectrum relative to
that in the magnetic field, $R=u_W/u_B$, we find that for the model
source parameters fit in \S3, $\lambda$ and $t_s$ depend on values
assumed for the turbulence index $q$ and $u_W$. For stochastic
acceleration by whistler waves to be relevant we require $\lambda
\ll L$ and $t_s\approx 50$ s. These requirements can be met when
$q\sim 2$ and $R\approx 10^{-4}$. With $t_s\propto
(\beta\gamma)^{q-2}/\beta$ and $q\sim 2$, $\lambda$ is essentially
independent of energy and $t_s$ depends only weakly on energy. We
therefore conclude that if stochastic processes in a turbulent
spectrum of whistler waves are relevant, then the spectrum must be
relatively flat (e.g., a Kolmogorov-like spectrum with $q=5/3$; cf.
Petrosian \& Liu 2004).  With the estimated turbulence level and
spectrum, the characteristic acceleration time for MeV electrons is
very short, a fraction of a second. This seems to contradict the
progression of delays up to $\sim 6$ s observed between the onset of
the HXR L band emission and the radio emission. The apparent
contradiction can perhaps be understood in terms of the evolution
time of the turbulence spectrum. But until the specific turbulent
wave mode is firmly established, and the source of the turbulence
identified, it is not possible to draw any firm conclusions in this
regard.

\section{Summary}

The flare on 2001 Oct 24 was well-observed at radio and X-ray
wavelengths. It was one of the most X-ray poor events jointly
observed by the HXT and the NoRP. Fortuitously, comprehensive X-ray and radio observations of the event. The spatial morphology of the source is consistent with a simple coronal magnetic loop. The
HXR and radio light curves revealed a progressive delay between
the onset of HXT L band emission (13.9-22.7 keV) and the onset of
NoRP/OVSA emission, the delay increasing with frequency for
$\nu\gtrsim 12$ GHz. In contrast, the time of the maximum flux
density was progressively delayed with decreasing frequency. The
radio emission decay was independent of frequency.

We have modeled the flare source in terms of an admixture of
nonthermal electrons and a dense thermal plasma. We suggest that the
paucity of EUV, SXR, and HXR emission as a result of a relatively
large mirror ratio in the magnetic loop in which the nonthermal
electrons are injected/accelerated and the high ambient plasma density, which rendered the source collisionally thick to electrons $\lesssim 30$ keV. The progressive delay in the
impulsive onset of radio emission with increasing frequency
($\nu\gtrsim 12$ GHz) is consistent with a progressive energization
of electrons from lower to higher energies during the impulsive rise
($\lesssim 10$ s). Electrons with energies of 10s of keV
deposit their energy in the coronal loop rather than at the foot points, heating the ambient
plasma. The systematic plasma heating with time causes the thermal free-free
absorption to decrease with time, thereby explaining the progressive delay of the
radio flux maximum with decreasing frequency. The radio decay time
is inconsistent with a trapping scenario. We suggest that the
transport of energetic electrons is instead determined by
wave-particle interactions. We suggest that the data are consistent with electron
acceleration and transport mediated by a relatively flat
(Kolmogorov) spectrum of whistler waves although the specifics remain an outstanding question.

\acknowledgements{We thank Hugh Hudson for comments on an early
draft of this paper and the referee for useful suggestions. We are
grateful to the Nobeyama Observatory for data from the NoRH and
NoRP. {\sl Yohkoh} was a mission of the Institute of Space and
Astronautical Sciences in Japan, in collaboration with the US and
UK. TRACE is a NASA Small Explorer and a mission of the
Stanford-Lockheed Institute for Space Research. We acknowledge use
of a magnetogram from the Kitt Peak National Observatory and an
H$\alpha$ filtergram from the Big Bear Solar Observatory. The
National Radio Astronomy Observatory is a facility of the National
Science Foundation operated under cooperative agreement by
Associated Universities, Inc. GDF acknowledges that this work was
supported in part by the Russian Foundation for Basic Research,
grants 06-02-16295a and 06-02-16859a. DEG acknowledges that this
work was supported in part by NSF grant AST-0607544 and NASA grant
NNG06GJ40G to the New Jersey Institute of Technology.}

\clearpage

% Figure 1 ----------------------------------------------------------------------------------
\begin{figure}
\plotone{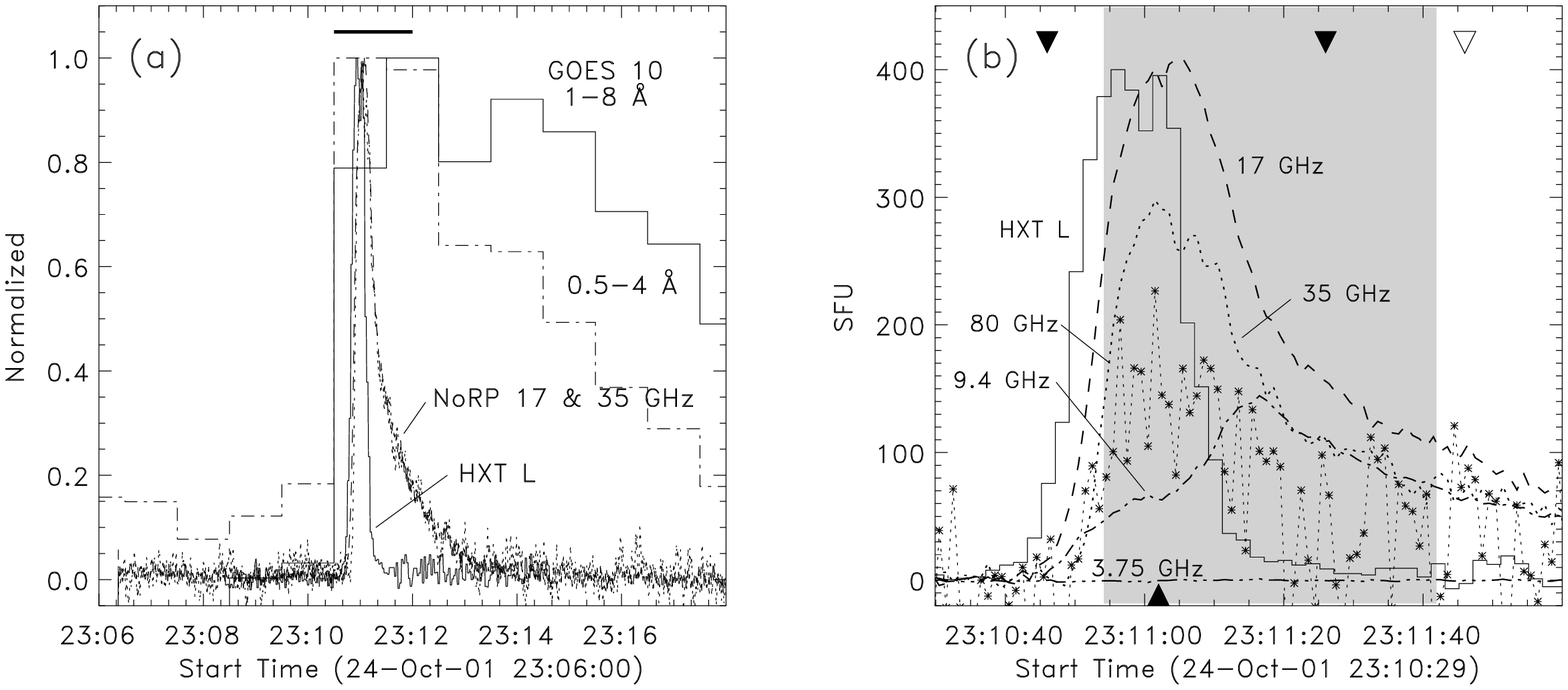} \caption{a) Overview of the radio and
X-ray emission from the flare that occurred at 23:11 UT on 2001
October 24. The {\sl Yohkoh} HXT L band (13.9-22.7 keV) photon
counts, the GOES 10 0.5-4\AA\ and 1-8\AA\ SXR fluxes averaged to 1
min, and the NoRP 17 and 35 GHz emissions are shown. Each trace is
normalized to its maximum. b) Detail of the NoRP and HXT L
observations, corresponding the time range indicated by the
horizontal bar at the top of panel (a). The radio flux densities are
indicated by the ordinate label. The HXT L band
background-subtracted counts (solid line; maximum of 6 cts sc$^{-1}$
s$^{-1}$) have been scaled for ease of comparison. The solid
inverted triangles indicate the times of the TRACE 171 \AA\ images.
The open inverted triangle indicates the time the single SXT full
disk image (AlMg filter). The shaded area indicates the time range
for which model fits to the radio spectroscopic data were obtained
(see \S3.2).}
\end{figure}

\clearpage

% Figure 2 -----------------------------------------------------------------------------------
\begin{figure}
\plotone{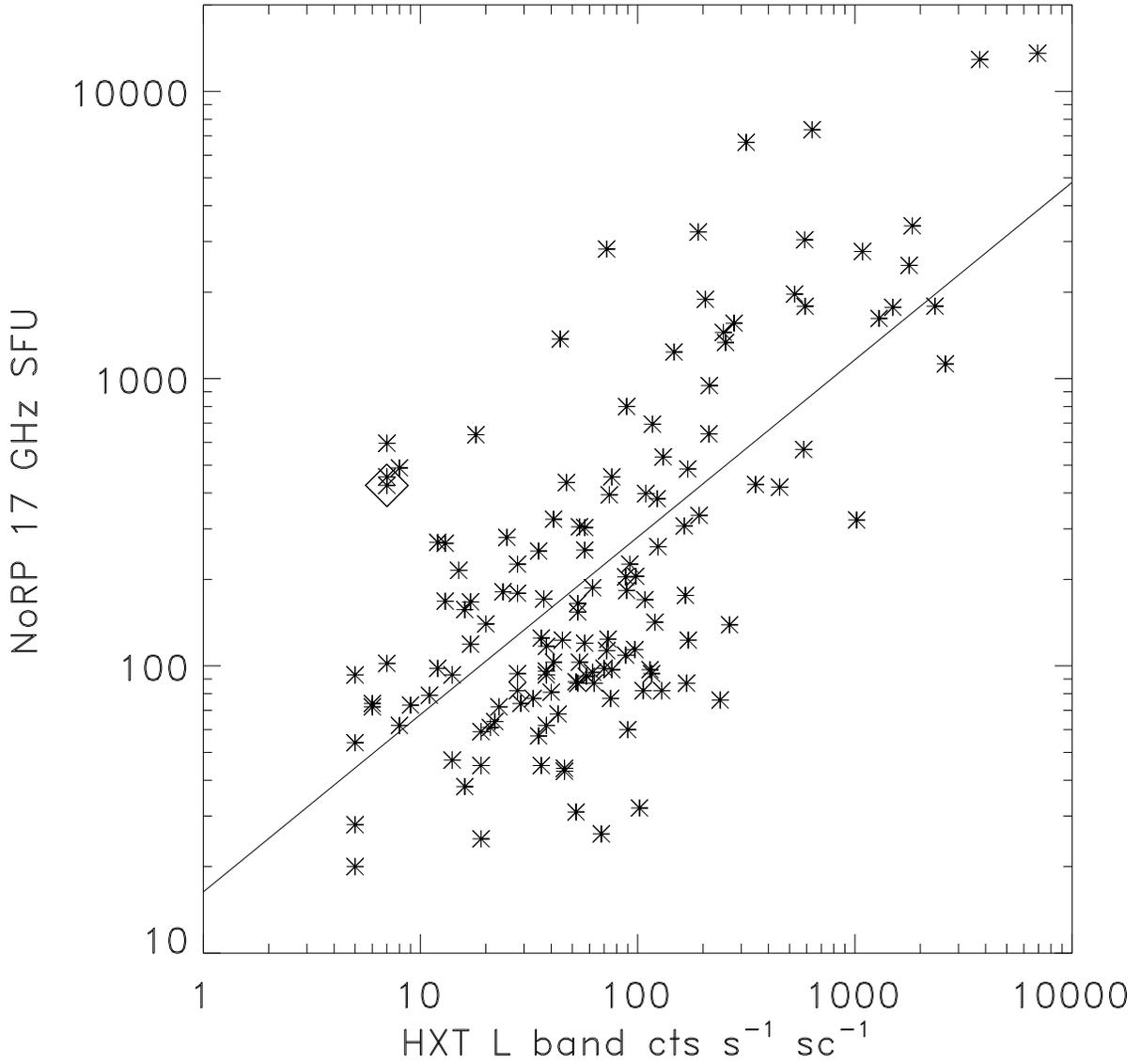} \caption{Plot of the NoRP 17 GHz flux
versus the {\sl Yohkoh} HXT L band (13.9-22.7 keV) count rate for
the 135 flares coincident to 10 min and observed jointly by the two
instruments during the lifetime of the {\sl Yohkoh} mission. Only
events above 20 SFU at 17 GHz and 4 cts s$^{-1}$ sc$^{-1}$ in the
HXT L band are shown. The flare of 2001 October 24 is indicated by
the diamond symbol.}
\end{figure}

\clearpage

% Figure 3 ----------------------------------------------------------------------------------
\begin{figure}
\plotone{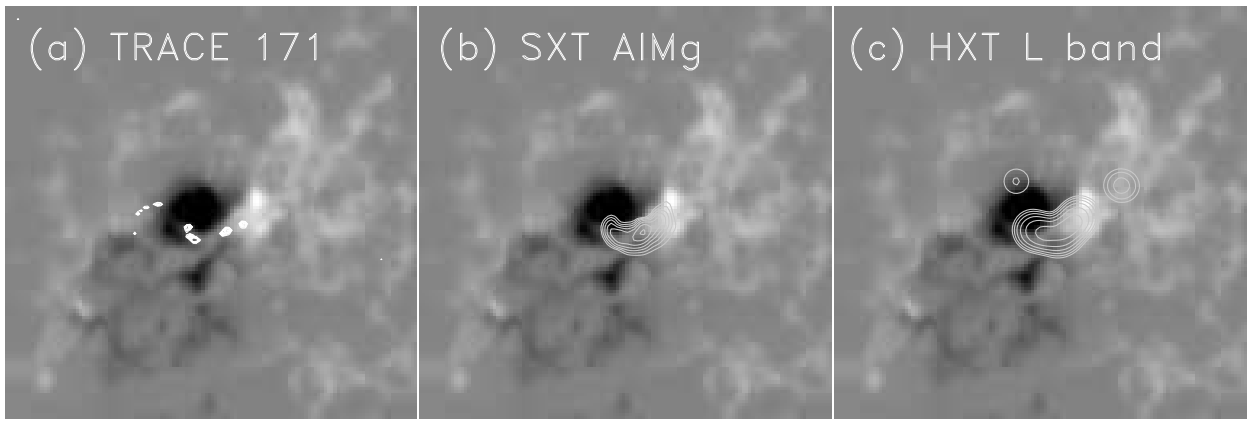} \caption{ A comparison of TRACE, SXT, and
HXT maps and a KPNO magnetogram. a) A difference map formed from
TRACE 171 \AA\ images obtained at 23:10:05 UT and 23:11:26 UT; b)
The SXT image detail obtained from the full disk image at 23:11:46
UT with the AlMg filter. b) The HXT L band (13.9-22.7 keV) map using
counts accumulated for the duration of the impulsive spike.}
\end{figure}

\clearpage

% Figure 4 ----------------------------------------------------------------------------------
\begin{figure}
\plotone{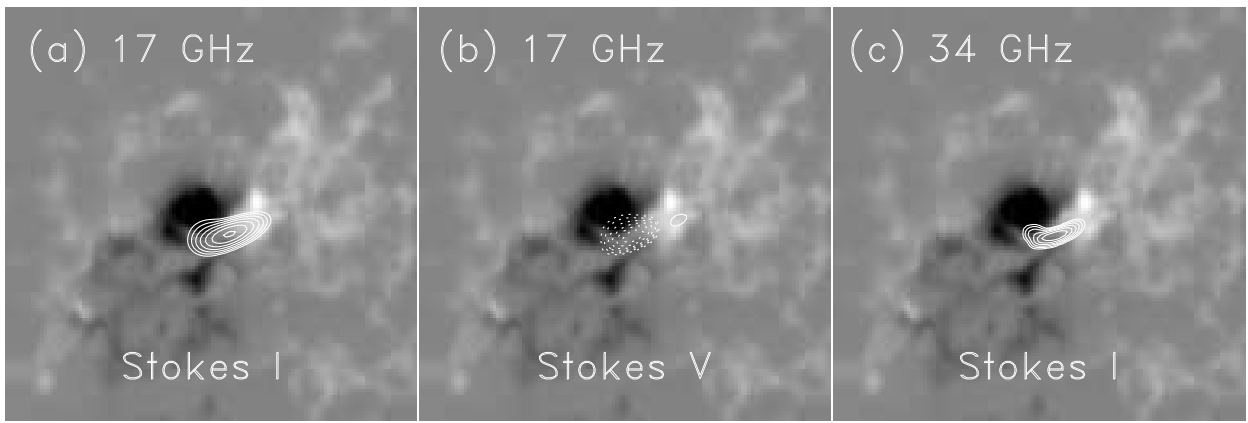} \caption{A comparison of the NoRH maps
with the KPNO magnetogram. a) The map of the 17 GHz total intensity
(Stokes I parameter) at the time of 17 GHz flux maximum (23:11:10
UT). The maximum brightness temperature is $4.6\times 10^7$ K. The
contour levels are at intervals of $\sqrt{2}$ from 12\% to 96\% of
the maximum. b) The map of the 17 GHz circularly polarized intensity
(Stokes V parameter) at the time of the 17 GHz total intensity
maximum. The contour levels are again 12\% to 95\% of the maximum at
intervals of $\sqrt(2)$; c) The map of the 34 GHz total intensity at
the time of the 17 GH maximum. The maximum brightness temperature is
$1.8\times 10^7$ K. The contour levels are again 12\% to 95\% of the
maximum brightness.}
\end{figure}

\clearpage

% Figure 5 ----------------------------------------------------------------------------------
\begin{figure}
\plotone{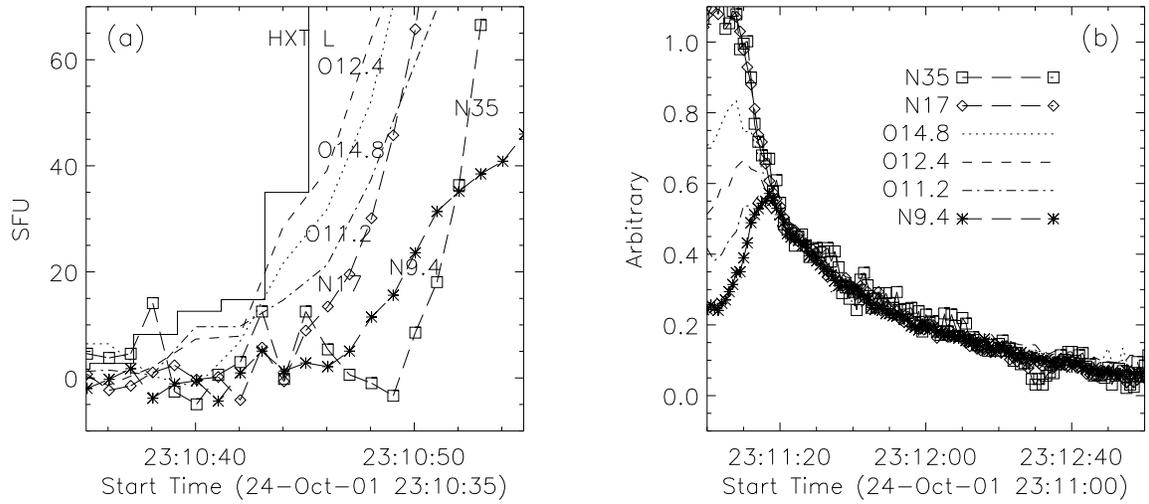} \caption{a) Detail of the impulsive rise
of of the burst comparing the HXT L band and selected NoRP and OVSA
frequencies. b) Detail of the decay phase. The NoRP and OVSA
frequencies shown in panel (a) have been scaled to emphasize that
the decay rate of each is essentially identical after 23:11:20 UT.
The delay in the flux maximum as a function of decreasing frequency
is also evident.}
\end{figure}

\clearpage

% Figure 6 ----------------------------------------------------------------------------------
\begin{figure}
\plotone{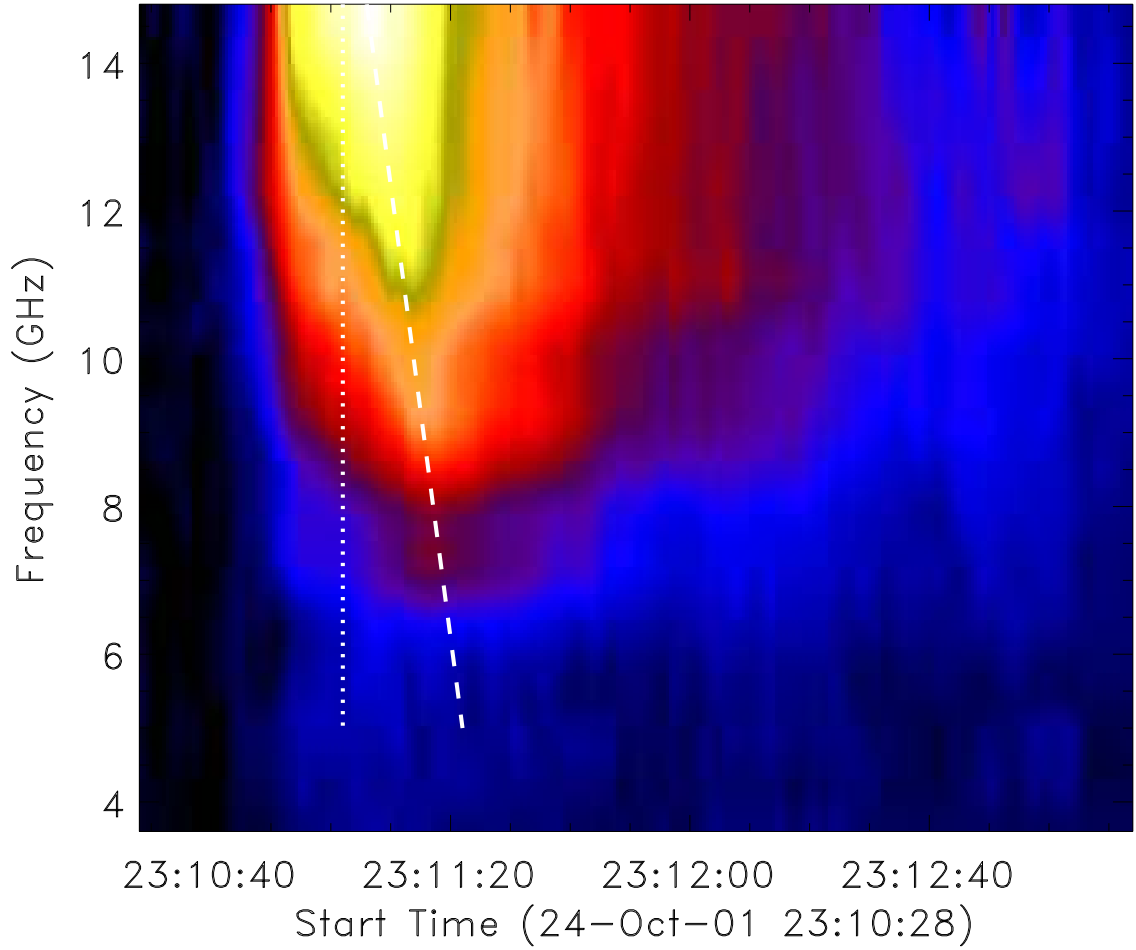} \caption{A dynamic spectrum of the OVSA
data. The dashed line emphasizes the increasing time of the flux
maximum as a function of decreasing frequency. The dotted vertical
line indicates the time of the 35 GHz flux maximum.}
\end{figure}

\clearpage

% Figure 7 ----------------------------------------------------------------------------------
\begin{figure}
\plotone{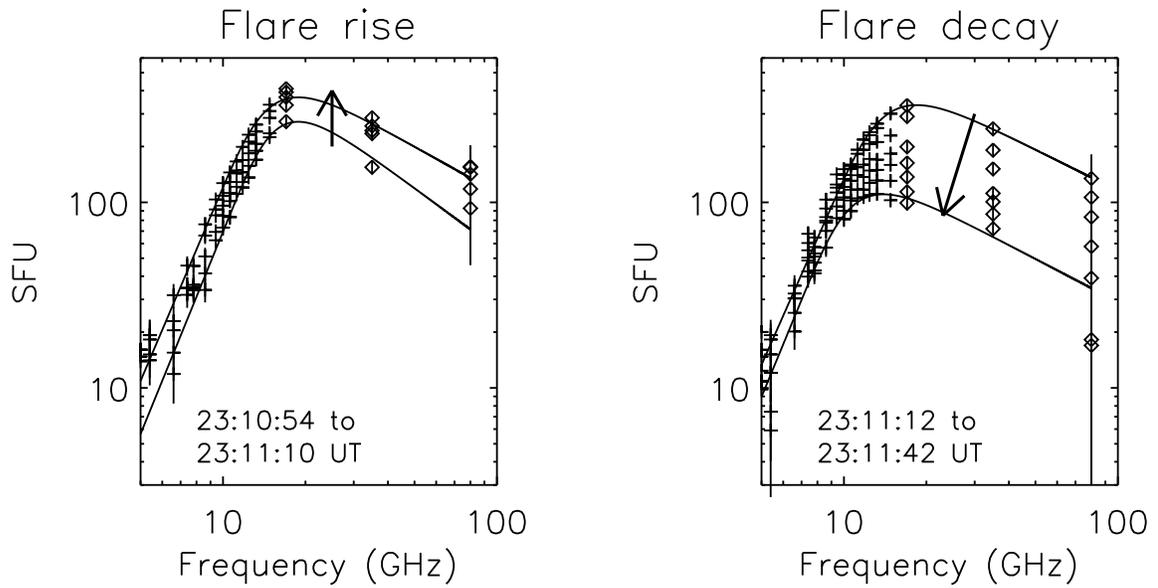} \caption{Stack plots of composite spectra
shown using OVSA data (asterisks) and NoRP data (squares) during (a)
the flare rise phase, during which spectra are shown at 4~s
intervals; b) the flare decay, during which spectra are shown at 6~s
intervals.}
\end{figure}

\clearpage

% Figure 8 ----------------------------------------------------------------------------------
\begin{figure}
\epsscale{1.0} \plotone{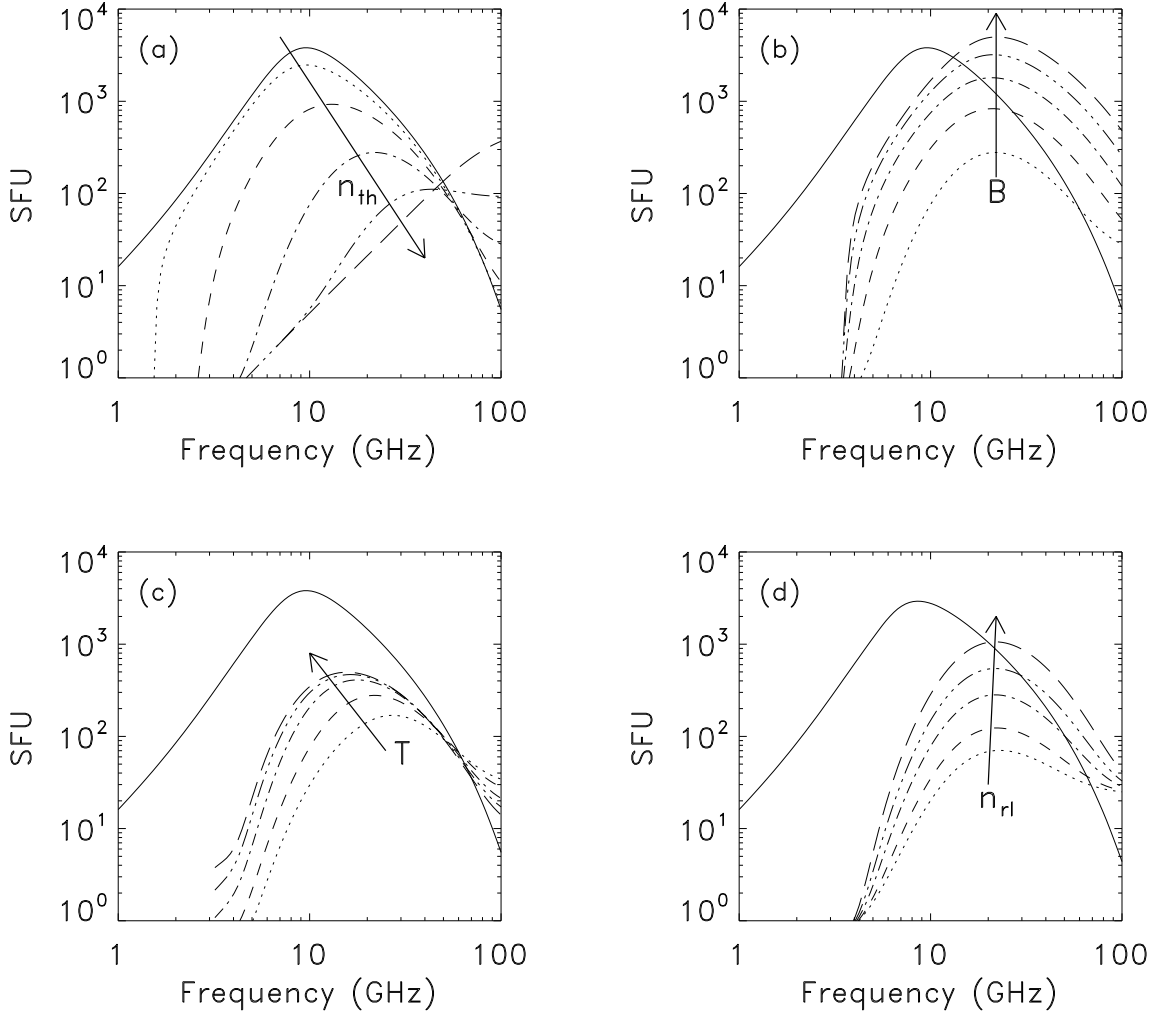} \caption{Gyrosynchrotron
spectra computed to illustrate parameter dependencies. In all cases,
the solid line represents gyrosynchrotron emission from electrons
{\sl in vacuo} and by parameters given in the text.  a) variation of
the spectrum when a background thermal plasma with a temperature
$T=2\times 10^6$ K is present: dotted line: $n_{th}=2\times 10^{10}$
cm$^{-3}$; dashed line: $n_{th}=5\times 10^{10}$ cm$^{-3}$;
dot-dash: $n_{th}=2\times 10^{11}$ cm$^{-3}$; dot-dot-dot-dash:
$n_{th}=2\times 10^{11}$ cm$^{-3}$; long-dash: $n_{th}=5\times
10^{11}$ cm$^{-3}$; b) same, except $n_{th}=10^{11}$ cm$^{-3}$,
$n_{rl}=5\times 10^6$ cm$^{-3}$, $T=2\times 10^6$ K, and $B=150,
200, 250, 300, 350$ G, respectively; c) same as (b) except $B=150$ G
and $T=10^6, 2\times10^6, 5\times 10^6, 10^7$, and $2\times 10^7$ K,
respectively; (d) same as (c) except $T=2\times 10^6$ and
$n_{rl}=10^6, 2\times 10^6, 5\times 10^6, 10^7, 2\times 10^7$
cm$^{-3}$, respectively. }
\end{figure}

\clearpage

% Figure 9 ----------------------------------------------------------------------------------
\begin{figure}
\epsscale{1.0} \plotone{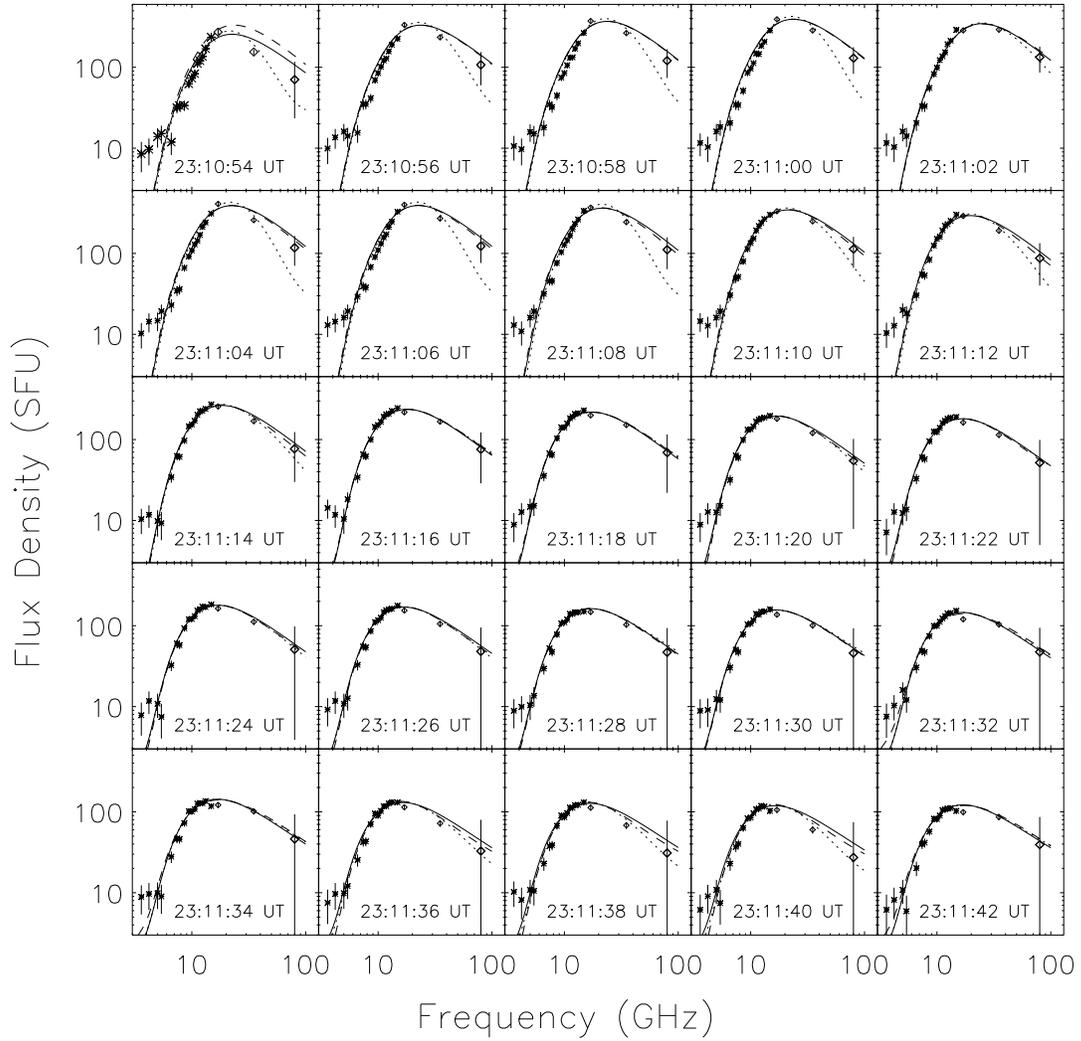} \caption{Time sequence of
radio spectra and model fits are shown with two-parameter (solid
line) and three-parameter (dashed lines) fits. See the text for
details. }
\end{figure}

\clearpage

% Figure 10 ----------------------------------------------------------------------------------
\begin{figure}
\plotone{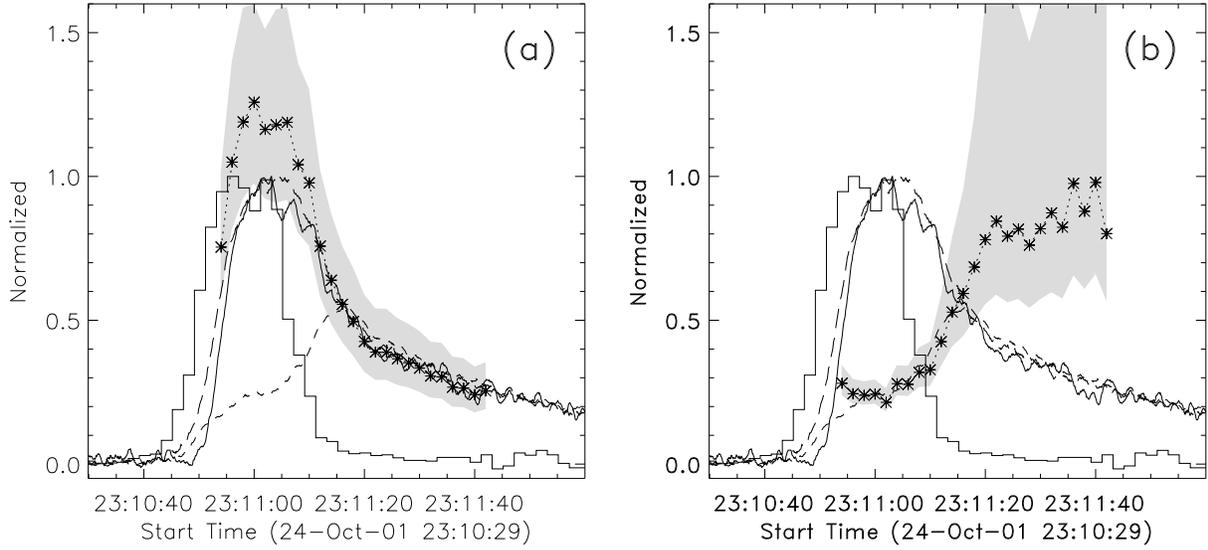} \caption{a) The variation of $n_{rl}$
with time in comparison to the HXT L band  (thin solid  line) and
radio emission at 9.4 (short dashes), 17 (long dashes), and 35 GHz
(heavy solid line). The HXR and radio data have been normalized to
their respective maxima. The dotted line connecting asterisks shows
the variation of the fitted value of $n_{rl}$ (divided by $10^7$
cm$^{-3}$) for an assumed magnetic field strength of 165 G in the
source. The shaded area indicates the range of 
$n_{rl}$ for assumed magnetic field strengths of 150 and 180 G. b) The same as panel (a), except the asterisks show
the variation of the fitted value of the temperature $T$ (divided by
$10^7$ K) of the ambient plasma for an assumed magnetic field
strength of 165 G in the source. The shaded area indicates the
temperatures when the magnetic field is 150 and 180 G.    }
\end{figure}

\end{document}